\documentclass[a4paper,12pt]{article}

\usepackage{epsfig}
\usepackage{amsfonts,eucal}

\usepackage[T2A]{fontenc}
\usepackage[cp1251]{inputenc}

\begin{document}

\title{A quasispecies continuous contact model in a critical regime
\thanks{The work is partially supported by SFB 701 (Universitat Bielefeld); the research of Sergey Pirogov and Elena Zhizhina was supported by the Russian Foundation for Sciences (project № 14-50-00150)} }

\author{Yuri Kondratiev\thanks{Fakultat fur
Mathematik, Universitat Bielefeld, 33615 Bielefeld, Germany
(kondrat@math.uni-bielefeld.de).} \and Sergey Pirogov\thanks{
Institute for Information Transmission Problems, Moscow, Russia
(s.a.pirogov@bk.ru).} \and Elena Zhizhina\thanks{Institute for
Information Transmission Problems, Moscow, Russia (ejj@iitp.ru).} }

\date{}

\maketitle

\begin{abstract}
We study a new non-equilibrium dynamical model: a marked continuous contact model in $d$-dimensional
space ($d \ge 3$). We prove that for certain values of rates (the critical regime) this
system has the one-parameter family of invariant measures labelled
by the spatial density of particles. Then we prove that the process starting from the marked Poisson measure converges to one of these invariant measures. In contrast with the continuous contact model studied earlier in \cite{KKP}, now the spatial particle density is not a conserved quantity.
\\

\end{abstract}

\section{Introduction}

In this paper we study a marked continuous contact model in $d$-dimensional
space ($d \ge 3$). This model can be considered as a
special case of birth-and-death processes in the continuum,
\cite{KKP, KS}, and it is inspired by the concept of quasispecies in population genetics, \cite{ES, N}.
The phase space of such processes is the space $
\Gamma = \Gamma (R^d \times S)$ of locally finite marked
configurations in $R^d$ with marks $s \in S$ from a compact metric
space $S$. Our purpose here is to describe various stationary
regimes and to specify relations between solutions of the Cauchy
problem and these stationary regimes.

The analysis of the model is based on the concept of statistical dynamics, see \cite{FKK}. Instead of the construction of the stochastic dynamics as a Markov process on the configuration space, we use here the formal generator of the model for the derivation of the hierarchical chain for the correlation functions similar to the BBGKY hierarchy for Hamiltonian dynamics. In the general framework of \cite{FKK} we use the hierarchy of equations for time-dependent correlation functions to describe the Markov dynamics of our system, c.f. also \cite{KKP}. The proof of our results is based on the techniques from \cite{KKP} combined with the Krein-Rutman theorem. To study the new situation (compact marks) we had to modify and generalize some steps of the proof presented in \cite{KKP}.

With biological point of view, the stochastic system under study can
be considered as a model of an asexual reproduction under mutations
and selections, where an individual at the point $u \in R^d$ with
the genome $s \in S$ produces an offspring distributed in the
coordinate space and in the genome space with the rate $\alpha(u-v)
Q(s,s')$. The function $Q(s,s')$ is said to be the mutation kernel.
Moreover, since mortality rates in our model can depend on genomes,
then selection rules are also included in the evolution under
consideration.

As in \cite{KKP} we prove in this paper the existence of the
stationary distributions for the marked contact model in
$d$-dimensional continuous space, $d \ge 3$, including 
the case of species dependent mortality, see Theorems 1-2 below.
Invariant distributions form the
one-parameter family parametrized by the spatial density of
particles. The invariant distributions are not Poisson and the
marks of neighboring particles are not independent random variables.  The
origin of this dependence is the existence of recent common
ancestors for spatially close individuals. In contrast to
\cite{KKP}, the spatial density for considered system is not
a conserved quantity. So the asymptotic value of the density can
differ from its initial value, see Remark 1 below.

\section{Main results.}

\subsection{Homogeneous mortality rates}

We consider a quasispecies contact model on $M = R^d \times S$,
where $d \ge 3$ and $S$ is a compact metric space. A heuristic
description of the process is given by a formal generator:
\begin{equation}\label{generator}
(L F)(\gamma) = \sum_{ x \in \gamma}(F(\gamma \backslash x) - F(
\gamma)) + \kappa \int_M \sum_{y \in \gamma}  a(x,y) (F(\gamma \cup
x ) - F(\gamma)) dZ(x),
\end{equation}
where $dZ = d\lambda d\nu $ is a product of the Lebesgue measure $
\lambda$ on $R^d$ and some finite Borel measure $\nu$ on $S$ with
$supp \ \nu =S$, and $F(\gamma)$ is defined by (\ref{FF}). Below we will construct
the operator $\hat L^{\ast}$ describing the evolution of the correlation functions
(the BBGKY type hierarchy equations). In our case these equations have the form (\ref{59})-(\ref{korf}).

Here $b(x, \gamma) = \kappa  \sum_{y \in \gamma} a(x,y)$ are birth
rates related to the contact model, and $m(x, \gamma) \equiv 1 $ are
death (mortality) rates. We take $a(x,y)$ in the following form:
\begin{equation}\label{a}
a(x,y) = \alpha (\tau(x) - \tau(y)) \, Q(\sigma(x), \sigma(y)),
\end{equation}
$\tau$ and $\sigma$ are projections of $M$ on $R^d$ and $S$
respectively, $\alpha(u) \ge 0$ is a function on $R^d$ such that
\begin{equation}\label{2a}
\int_{R^d} \alpha(u) du \ = \ 1,
\end{equation}
\begin{equation}\label{2c}
 \int_{R^d} |u|^2 \alpha(u) du \ < \ \infty,
\end{equation}
the covariance matrix $C$
\begin{equation}\label{2d}
C_{jk} \ = \ \int_{R^d} u_j u_k \alpha(u) du \ - \ m_j m_k, \quad
m_j \ = \ \int_{R^d} u_j \alpha(u) du,
\end{equation}
is non-degenerate and
\begin{equation}\label{2b}
 \hat \alpha (p) \ = \ \int_{R^d} e^{i(p,u)} \alpha(u) du \in
L^1(R^d).
\end{equation}
It follows in particular that $|\hat \alpha (p)|<1$ for all $p \neq
0$.

We suppose that the function $Q$ on $S \times S$ is continuous on
$S \times S$ (and so bounded) and strictly positive. Then the Krein-Rutman theorem
\cite{KR} implies that there are a positive number $r>0$ and a
strictly positive continuous function $q(s)$ on $S$, such that $Qq \
= \ rq$ for the integral operator
\begin{equation}\label{Q}
(Qh)(s) \ = \ \int_S Q(s, s') h(s') d\nu(s'),
\end{equation}
and the spectrum of $Q$, except $r$, which is a discrete spectrum
accumulated to 0, is contained in the open disk $\{ z: \ |z|<r \}
\subset \mathbb{C}$. (Here we consider the spectrum of the integral
operator (\ref{Q}) in the Banach space of continuous functions
$C(S)$). This "rest spectrum" is the spectrum of $Q$ on the subspace
"biorthogonal to $q$", i.e. on the subspace of the functions $h(s)$
such that
$$
\int_S h(s) \ \tilde q(s) \  d\nu(s)\  =  \ 0.
$$
Here $\tilde q(s)$ is the stricily positive eigenfunction of the
adjoint operator $Q^{\star} (s,s') = Q(s', s)$. We take $\kappa = \kappa_{cr}
= r^{-1}$ and now including $\kappa_{cr}$ in $Q$ we shall suppose
that $r=1$, i.e. $Qq=q$. So the "renormalized critical value of
$\kappa$" equals 1 and we omit $\kappa$ in (\ref{generator}) in what
follows. We also normalize the function $q$ by the condition
\begin{equation}\label{norm}
\int_S q(s) d\nu(s) \ = \ 1.
\end{equation}

Note that the existence problem for Markov processes in $\Gamma$ for
general birth and death rates is an essentially open problem. An
alternative way of studying the evolution of the system is to
consider the corresponding statistical dynamics. The latter means
that instead of a time evolution of configurations we consider a
time evolution of initial states (distributions), i.e. solutions of
the corresponding forward Kolmogorov (Fokker-Planck) equation, see
(\cite{FKK, KKZ}) for details.

We should remind basic notations and constructions to derive time
evolution equations on correlation functions of the considered
model. Let ${\cal B}(M)$ be the family of all Borel sets in $M = R^d
\times S$, and ${\cal B}_b (M) \subset {\cal B}(M)$ denotes the
family of all bounded sets from ${\cal B}(M)$. The configuration
space $\Gamma (M)$ consists of all locally finite subsets of $M$:
$$
\Gamma \ = \  \Gamma (M) \ =  \ \{ \gamma \subset M: \ |\gamma \cap
\Lambda| < \infty \; \mbox{ for all } \; \Lambda \in {\cal B}_b(M)
\}.
$$
Together with the configuration space $\Gamma(M)$ we define the
space of finite configurations
$$
\Gamma_0 \  = \  \Gamma_0 (M) \ = \ \bigsqcup_{n \in N \cup \{0 \} }
\ \Gamma_0^{(n)},
$$
where $\Gamma_0^{(n)}$ is the space of $n$-point configurations:
$$
\Gamma_0^{(n)} \ = \ \{ \eta \subset M : \ |\eta| = |\tau(\eta)| = n
\}.
$$
We denote the set of bounded measurable functions with bounded
support by $B_{bs}(\Gamma_0)$, and the set of cylinder functions on
$\Gamma$ by ${\cal F}_{cyl}(\Gamma)$. Each $F \in {\cal
F}_{cyl}(\Gamma)$ is characterized by the following relation:
$F(\gamma) = F(\gamma_{\Lambda})$ for some $\Lambda \in {\cal B}_b
(M)$.

Next we define a mapping from $B_{bs}(\Gamma_0)$ into ${\cal
F}_{cyl}(\Gamma)$ as follows:
\begin{equation}\label{FF}
(K \ G)(\gamma) \ = \ \sum_{\eta \subset \gamma} G(\eta), \quad
\gamma \in \Gamma, \; \eta \in \Gamma_0,
\end{equation}
where the summation is taken over all finite subconfigurations $\eta
\in \Gamma_0$ of the infinite configuration $\gamma \in \Gamma$, see
i.g. \cite{KKP} for details. This mapping is called K-transform.
\\

{\bf Proposition 1. }  {\it  The operator $\hat L = K^{-1}LK$ (the
image of $L$ under the K-transform) on functions   $G \in
B_{bs}(\Gamma_0)$ has the following form:}
\begin{equation}\label{prop1}
(\hat L G)(\eta) \ = \ - |\tau(\eta)| G(\eta)  \  + \  \int_M
\sum_{y \in \eta} a(x,y) G((\eta \backslash y) \cup x) dx \  +
\end{equation}
$$
\int_M \sum_{y \in \eta} a(x,y) G(\eta \cup x) dx.
$$

The derivation of the formula (\ref{prop1}) is the same as in
\cite{KKP}.
\\

Denote by ${\cal M}^1_{fm}(\Gamma)$ the set of all probability
measures $\mu$ which have finite local moments of all orders, i.e.
$$
\int_{\Gamma} |\gamma_{\Lambda}|^n \ \mu (d \gamma) \ < \ \infty
$$
for  all $\Lambda \in {\cal B}_b(M)$ and $n \in N.$ If a measure
$\mu \in {\cal M}^1_{fm}(\Gamma)$ is locally absolutely continuous
with respect to the Poisson measure (associated with the measure
$dZ$), then there exists the corresponding system of the correlation
functions $k_{\mu}^{(n)}$ of the measure $\mu$, well known in
statistical physics, see e.g. \cite{R}.

Let $\{ \mu_t \}_{t \ge 0} \subset {\cal M}_{fm}^1
(\Gamma)$ be the evolution of states described by the dual Kolmogorov equation with
the adjoint operator $L^{\ast}$. Then the evolution of the
corresponding system of correlation functions is defined by the
duality equation
$$
\langle \hat L G, k \rangle \ = \  \langle G, \hat L^{\ast} k
\rangle, \quad  G \in B_{bs}(\Gamma_0),
$$
where the operator $\hat L$ is defined by (\ref{prop1}).
Using the representation (\ref{prop1}) we define the operator $\hat
L^{\ast}$ adjoint to the operator $\hat L$ and obtain the following
system of equations for correlation functions in a recurrent form:
\begin{equation}\label{59}
\frac{\partial k^{(n)}}{\partial t} \ = \ \hat L_n^{\ast} k^{(n)} \
+ \ f^{(n)}, \quad n\ge 1; \quad f^{(1)}=0,
\end{equation}
which is the main object for study in this paper. Here $f^{(n)}$ is
a function on $M^n$ defined as
\begin{equation}\label{f}
f^{(n)}(x_1, \ldots, x_n) \ = \ \sum_{i=1}^n  k^{(n-1)}(x_1,
\ldots,\check{x_i}, \ldots, x_n) \sum_{j\neq i}^n a(x_i, x_j), \; n
\ge 2,
\end{equation}
$f^{(1)} \equiv 0$. The operator $\hat L^{\ast}_n, \; n \ge 1, $ is
defined as:
$$
\hat L^{\ast}_n k^{(n)}(x_1, \ldots, x_n) \ = \ - n k^{(n)}(x_1,
\ldots, x_n) +
$$
\begin{equation}\label{korf}
\sum_{i=1}^n \int_M a(x_i, y) k^{(n)}(x_1, \ldots, x_{i-1}, y,
x_{i+1}, \ldots, x_n) dy.
\end{equation}
As follows from (\ref{a}) the operators $\hat L^{\ast}_n$ are bounded.

We take the initial (for $t=0$) data
\begin{equation}\label{60}
k^{(n)}(t=0, \varrho; x_1, \ldots, x_n) \ = \ \varrho^n \
\prod_{i=1}^{n} q(\sigma(x_i)).
\end{equation}
corresponding to the marked Poisson point field with the intensity
$\varrho$ and the distribution of marks $q(s) d\nu(s)$ meeting
(\ref{norm}), where $q(s)$ is the eigenfunction of $Q$: $Qq=q$. In fact, we consider in the paper only the case when
the initial measures associated to these correlation functions is a marked Poisson measure with the intensity $\rho q(s) d\lambda d\nu(s)$, see also a general form (\ref{93}) for the initial data in Remark 1 below.

Invariant measures of the contact process (if exist!) are described
in terms of correlation functions $k^{(n)}$ on $M^n$ as a positive
solutions of the following system:
\begin{equation}\label{Last}
\hat L^{\ast}_n k^{(n)} + f^{(n)}=0, \quad n \ge 1, \quad
k^{(0)}\equiv 1,
\end{equation}
where $\hat L_n^{\ast}, \, f^{(n)}$ are defined as in (\ref{f}) -
(\ref{korf}).

Consider the operator $\hat L_n^{\ast}$ as an operator on the space
$$
X_n = C \left( S^n, \ L^{\infty}_{inv}((R^n)^d) \right),
$$
where $L^{\infty}_{inv}$ consists of the bounded translation
invariant functions $\varphi (w_1, \ldots, w_n)$ of $n$ variables:
$$
\varphi (w_1+a, \ldots, w_n+a) = \varphi (w_1, \ldots, w_n), \quad
 w_i = \tau (x_i) \in R^d.
$$
In this section we prove the existence of the solution $k^{(n)} \in
X_n, \, n \ge 1$ of the system (\ref{Last}), such that $k^{(n)}$
have a specified asymptotics when $|\tau(x_i)-\tau(x_j)|\to\infty$
for all $i \neq j$. We also prove a strong convergence of the
solutions of the Cauchy problem (\ref{59}) - (\ref{60}) to the
solution of the system (\ref{Last}) of stationary (time-independent)
equations.
\\

{\bf Theorem 1.} {\it I. Let the birth kernel $a(x,y)$ of the
contact model meet conditions (\ref{a})-(\ref{Q}), and $\kappa = \kappa_{cr}
= r^{-1}$.

Then for any positive constant $\varrho \in R_+$ there exists a
unique probability measure $\mu^{\varrho}$ such that its system of
correlation functions $\{k^{(n)}_\varrho \} $ is translation
invariant, solves (\ref{Last}), satisfies the following condition
\begin{equation}\label{Th1}
| k^{(n)}_\varrho (x_1, \ldots, x_n) \ - \ \varrho^n \prod_{i=1}^{n}
q(\sigma(x_i))| \ \to \ 0,
\end{equation}
when  $|\tau(x_i) - \tau(x_j)| \to \infty$ for all $i \neq j$, and
satisfies the following estimate
\begin{equation}\label{estimate}
k^{(n)}_\varrho (x_1, \ldots, x_n) \  \le \  D \ C^n (n!)^2
\prod_{i=1}^n q(\sigma(x_i)) \quad \mbox{for any } \ x_1, \ldots,
x_n,
\end{equation}
for some positive constants $C=C(\varrho, Q, \alpha), \ D=D(\varrho, Q, \alpha)$. Here
$q(s)$ is the normalized eigenfunction of $Q$. Moreover, the first
correlation function $k^{(1)}_\varrho(x)$ of $\mu^{\varrho}$ is
exactly $\varrho \ q(s)$.

II. For any $n \ge 1$ the solution $k^{(n)}(t)$ of the Cauchy
problem (\ref{59}) - (\ref{60}) converges to the solution
$k_\varrho^{(n)}$ (\ref{Th1}) of the system (\ref{Last}) of
stationary (time-independent) equations as  $t \to \infty$:
\begin{equation}\label{Th1-2}
\| k^{(n)}(t) \ - \ k_\varrho^{(n)} \|_{X_n} \ \to \ 0,
\end{equation}
where $X_n = C \left( S^n, \ L^{\infty}_{inv}((R^n)^d) \right)$.
}

\subsection{Species dependent mortality rates}

Analogous results are valid in the case when mortality rates depends
on $\sigma(x)$:
\begin{equation}\label{generatorS}
(\tilde L F)(\gamma) \ = \  \sum_{ x \in \gamma} m(\sigma(x)) \
(F(\gamma \backslash x) - F( \gamma)) \ + \ \kappa \int_M \sum_{y
\in \gamma} a(x,y) (F(\gamma \cup x ) - F(\gamma)) dx,
\end{equation}
with
$$
a(x,y) = \alpha (\tau(x) - \tau(y)) \, Q(\sigma(x), \sigma(y)),
$$
$\tau$ and $\sigma$ are projections of $M$ on $R^d$ and $S$
respectively, $m(\sigma(x))>0$. In this case using the Krein-Rutman
theorem for the integral operator $\tilde Q$ with the kernel
$$
\tilde Q(s, s') \ = \ \frac{Q(s,s')}{m(s)}
$$
we get the existence of the maximal eigenvalue $\tilde r>0$ and the
corresponding maximal positive eigenfunction $g(s)>0$ for the
operator $\tilde Q$.

Correlation functions for the invariant measure in this case can be
constructed as a solution of the system of equations
\begin{equation}\label{LastS}
\tilde L^{\ast}_n \tilde k^{(n)} + \tilde f^{(n)}=0, \quad n \ge 1,
\quad \tilde k^{(0)}\equiv 1,
\end{equation}
where
$$
\tilde f^{(n)}(x_1, \ldots, x_n) \ = \ \kappa \sum_{i=1}^n \tilde
k^{(n-1)}(x_1, \ldots,\check{x_i}, \ldots, x_n) \sum_{j\neq i}^n
a(x_i, x_j), \; n \ge 2,
$$
$\tilde f^{(1)} \equiv 0$,
$$
\tilde L^{\ast}_n \tilde k^{(n)}(x_1, \ldots, x_n) \ = \ -
\sum_{i=1}^n m(\sigma(x_i)) \ \tilde k^{(n)}(x_1, \ldots, x_n) \ +
$$
$$
\kappa \sum_{i=1}^n \int_M a(x_i, y) \tilde k^{(n)}(x_1, \ldots,
x_{i-1}, y, x_{i+1}, \ldots, x_n) dy.
$$
\\

{\bf Theorem 2. } {\it Let $m(s)>0, \; Q(s,s')>0, \ s,s' \in S$ are
continuous functions on $S$ and $S \times S$ respectively; $g(s)>0$
is the positive eigenfunction corresponding to the maximal
eigenvalue $\tilde r>0$ of the integral operator
$$
(\tilde Q h ) (x) \ = \ \int_S \tilde Q(s,s') h(s') d\nu(s'), \quad
\mbox{ with } \; \tilde Q(s, s') \ = \ \frac{Q(s,s')}{m(s)}.
$$
Let $\kappa_{cr} = {\tilde r}^{-1}$. Then for any positive constant
$\varrho \in R_+$ there exists a unique probability measure $\tilde
\mu^{\varrho}$ such that its system of correlation functions
$\{\tilde k^{(n)}_\varrho \} $ is translation invariant, solves
(\ref{LastS}), satisfies the following condition
\begin{equation}\label{99}
| \tilde k^{(n)}_\varrho (x_1, \ldots, x_n) \ - \ \varrho^n
\prod_{i=1}^{n} g(\sigma(x_i))| \ \to \ 0,
\end{equation}
when  $|\tau(x_i) - \tau(x_j)| \to \infty$ for all $i \neq j$, and
satisfies the following estimate
$$
\tilde k^{(n)}_\varrho (x_1, \ldots, x_n) \  \le \  D \ C^n (n!)^2
\prod_{i=1}^n g(\sigma(x_i)) \quad \mbox{for any } \ x_1, \ldots,
x_n,
$$
with positive constants $C, \ D$. Here $g(s)$ is the normalized
eigenfunction of the operator $\tilde Q$. The first correlation
function $\tilde k^{(1)}_\varrho(x)$ of $\tilde \mu^{\varrho}$ is
exactly $\varrho g(s)$.

Moreover, the solution of the Cauchy problem for the system of
equations
$$
\frac{\partial \tilde k^{(n)}}{\partial t} \ = \ \tilde L_n^{\ast}
\tilde k^{(n)} \ + \ \tilde f^{(n)}, \quad n\ge 1; \quad f^{(1)}=0,
$$
with the initial data
$$
\tilde k^{(n)}|_{t=0} (x_1, \ldots, x_n) \ = \ \varrho^n \
\prod_{i=1}^{n} g(\sigma(x_i)),
$$
converges to the system of the correlation functions $\{\tilde
k^{(n)}_\varrho \}$ defined by (\ref{99}). }

\section{The proof of Theorem 1. Stationary problem. }

In this section we prove the first part of Theorem 1 using the
induction in $n$. For $n=1$ in (\ref{Last}) we have
\begin{equation}\label{8}
-k^{(1)}(x) + \int_M a(x,y) k^{(1)}(y) dy = 0.
\end{equation}
As we construct a translation invariant field let us look for
$k^{(1)}(x)$ in the form
$$
k^{(1)}(x) \ = \ h(\sigma(x))
$$
Then (\ref{8}) can be rewritten as
\begin{equation}\label{10}
-h(s) + \int_S Q(s,s') h(s') d\nu(s') \ = \ 0,
\end{equation}
which means that
$$
h(s) \ = \ \varrho \ q(s),
$$
or
$$
k^{(1)}(x) \ = \ \varrho \ q(\sigma(x)),
$$
where $q(s)$ is the normalized eigenfunction of $Q$. From the normalization condition (\ref{norm}) it follows that
$\varrho$ can be interpreted as the spatial density of particles.
\\

As a warm-up let us solve the equation (\ref{Last}) for the special
case $n=2, \; S=\{0\}, \; m(0)=q(0)=1, \; Q(0,0)=1$. This means that
$M=R^d$ and that we have no marks. Then the equation for $k^{(2)}(x)$ is written as
\begin{equation}\label{13}
\hat L^{\ast}_2 k^{(2)} + f^{(2)}=0,
\end{equation}
with
\begin{equation}\label{14}
f^{(2)}(x_1, x_2) \ = \ \varrho( a(x_1, x_2) + a(x_2, x_1)) =
\varrho( \alpha(x_1 - x_2) + \alpha(x_2 - x_1)).
\end{equation}
Thus, the operator $\hat L^{\ast}_2 \ = \  L^{(1)} + L^{(2)}$, where
\begin{equation}\label{15}
L^{(1)} k^{(2)}(x_1, x_2) \ = \ \int_{R^d} \alpha(x_1 - y)
k^{(2)}(y, x_2) dy - k^{(2)}(x_1, x_2),
\end{equation}
and analogously
\begin{equation}\label{16}
L^{(2)} k^{(2)}(x_1, x_2) \ = \ \int_{R^d} \alpha(x_2 - y)
k^{(2)}(x_1, y) dy - k^{(2)}(x_1, x_2).
\end{equation}
Using translation invariant property we have:
$$
k^{(2)}(x_1, x_2) \ = \ k^{(2)}(x_1 - x_2).
$$
After the Fourier transform we can rewrite (\ref{13}) - (\ref{16})
as
\begin{equation}\label{18}
(\hat \alpha(p) + \hat \alpha (-p) -2) \ \hat k(p)  \ = \ -\varrho \
(\hat \alpha(p) + \hat \alpha (-p)).
\end{equation}
Therefore,
\begin{equation}\label{19}
\hat k(p)  \ = \ \varrho \ \frac{ \hat \alpha(p) + \hat \alpha (-p)}
{2 - \hat \alpha(p) - \hat \alpha (-p)} + A \delta(p),
\end{equation}
where $A$ is an arbitrary constant, and we will explain later how to
choose $A$ in the general case.

Expanding $\hat \alpha (p)$ in the Taylor series up to the second
order and using the conditions (\ref{2a}) - (\ref{2b}) on the
function $\alpha$ we see that $\hat k (p)$ has a
singularity $\sim |p|^{-2}$ at $p=0$ which is integrable if the dimension $d \ge 3$.
Thus there exist infinitely many translation invariant functions
$k^{(2)}(x_1 - x_2) \in L^{\infty} (R^d)$ satisfying equation
(\ref{13}).
\\

Now let us turn to the general case. If for any $n>1$ we succeeded to
solve the equation (\ref{Last}) and express $k^{(n)}$ through
$f^{(n)}$, then knowing the expression of $f^{(n)}$ through
$k^{(n-1)}$ via (\ref{f}), we would get the solution to the full system
(\ref{Last}). So we have to invert the operator $\hat L_n^{\ast}$,
and {\bf it is sufficient for us to do so on some class of
translation invariant functions}. The precise statement for $(\hat L_n^{\ast})^{-1} f^{(n)}$ will be
presented later, see formula (\ref{35}).

Remind that
\begin{equation}\label{20}
\hat L_n^{\ast} \ = \ \sum_{i=1}^n L^i,
\end{equation}
where
\begin{equation}\label{21}
L^{i} k^{(n)}(x_1, \ldots, x_n) \ =
\end{equation}
$$
\int_M a(x_i, y) k^{(n)}(x_1, \ldots, x_{i-1}, y, x_{i+1}, \ldots,
x_n) dy - k^{(n)}(x_1, \ldots, x_n)
$$
are bounded operators.
\\
{\bf Proposition 2.}{\it The operator $e^{t \hat L_n^{\ast}}$ is
monotone.} \\
{\bf Proof. } The monotonicity of the operator $e^{t \hat
L_n^{\ast}}$ follows from (\ref{20}) - (\ref{21}):
$$
e^{t \hat L_n^{\ast}} \ = \ \otimes_{i=1}^n  e^{t L^{i}}, \quad e^{t
L^{i}} \ = \ e^{-t} e^{t A^{i}},
$$
and the positivity of operators
$$
A^i k^{(n)} \ = \ \int_M a(x_i, y) k^{(n)} (x_1, \ldots, x_{i-1}, y,
x_{i+1}, \ldots, x_n) dy.
$$
$\Box$
\\

First consider the restriction of $\hat L_n^{\ast}$ to the invariant
subspace consisting of the functions of the form
$$
\varphi(\tau(x_1), \ldots, \tau(x_n)) \prod_{i=1}^n q(\sigma(x_i)),
\quad \mbox{ where } \; \varphi (w_1, \ldots, w_n) \in
L^{\infty}_{inv}((R^n)^d).
$$
The operator $\hat L_n^{\ast}$ acts on these functions as
\begin{equation}\label{24}
L_{n, max}  \ = \ \sum_{i=1}^n L^i_{max},
\end{equation}
where
\begin{equation}\label{25}
L^{i}_{max} \ \varphi (w_1, \ldots, w_n) \ \prod_{i=1}^n
q(\sigma(x_i)) \  =
\end{equation}
$$
\prod_{i=1}^n q(\sigma(x_i)) \left( \int_{R^d} \alpha(w_i-u)
\varphi(w_1, \ldots, w_{i-1}, u, w_{i+1}, \ldots, w_n) du - \varphi
(w_1, \ldots, w_n)\right)
$$
due to the equality $Qq=q$. Remind that
$\kappa_{cr}$ is "absorbed" in $Q$. Formula (\ref{25}) means that in
this case we have only spatial convolutions and no integration over
$S$. In the Fourier variables the operator $L_{n, max}$ acts as a
multiplication operator by the function
$$
\sum_{i=1}^n \hat \alpha (p_i) \ - \ n.
$$
To invert $L_{n, max}$ let us notice that if $\varphi (w_1, \ldots,
w_n)$ is a translation invariant function then its Fourier transform
has a form
$$
\hat \varphi (p_1, \ldots, p_n) \ \delta (p_1+ \ldots + p_n).
$$
On the subspace of the "momentum space" $(p_1, \ldots, p_n)$
specified by the equation $p_1+ \ldots + p_n = 0$ the function $
\frac{1}{\sum_{i=1}^n \hat \alpha (p_i) - n} $ has an integrable
singularity $\sim \frac{1}{|p|^2}$  at $p=0$. This property will be
crucial for inverting of the operator $L_{n, max}$ on a proper class
of functions.
\\

Next we will construct a solution of the system (\ref{Last}) satisfying (\ref{Th1}) and meeting the estimate
\begin{equation}\label{est}
k^{(n)} (x_1, \ldots, x_n) \ \le \ K_n \prod_{i=1}^n q(\sigma(x_i))
\end{equation}
where $K_n = D C^n (n!)^2$, $D,\ C$ are constants.

As follows from (\ref{f}), the function $f^{(n)}$ is the sum of
functions of the form
\begin{equation}\label{32}
f (x_1, \ldots, x_n) \ = \ k^{(n-1)} (x_1, \ldots,\check{x_i},
\ldots, x_n) \ a(x_i, x_j), \quad x_i \in M.
\end{equation}
Below we invert the operator $\hat L_n^{\ast}$ on the set of
functions of the form (\ref{32}), see (\ref{sumv}) below.

We put
\begin{equation}\label{35}
v^{(n)}_{i,j}  \ =  \ \int_0^{\infty} e^{t \hat L_n^{\ast}} f \ dt,
\end{equation}
where $f$ is a function of the form (\ref{32}), then
\begin{equation}\label{sumv}
k^{(n)} \  = \ \int_0^{\infty} e^{t \hat L_n^{\ast}} f^{(n)} \ dt \ = \
\sum_{i \neq j} v^{(n)}_{i,j}.
\end{equation}

We suppose by induction that
$$
k^{(n-1)} (x_1, \ldots, x_{n-1}) \ \le \ K_{n-1} \prod_{i=1}^{n-1}
q(\sigma(x_i)),
$$
then
\begin{equation}\label{34}
f(x_1, \ldots, x_n) \ \le \  K_{n-1} a(x_i, x_j) \prod_{l \neq i}
q(\sigma(x_l)).
\end{equation}

Since the function $q(s)$ is strictly positive on the compact $S$,
the following inequality holds:
\begin{equation}\label{aq}
a(x_i, x_j) \ \le \ c \ q(\sigma(x_i)) \ \alpha(\tau(x_i) -
\tau(x_j))
\end{equation}
with a constant $c$. Then using the monotonicity, identity $Qq=q$,
and inequality (\ref{aq}) we get from (\ref{34}), (\ref{20}) and
(\ref{24})
\begin{equation}\label{36}
e^{t \hat L_n^{\ast}} f  \ \le \  K_{n-1} \ e^{t \hat L_n^{\ast}}
a(x_i, x_j) \ \prod_{l \neq i} q(\sigma(x_l)) \ \le \ K_{n-1} \
e^{t ( L^i + L^j)} a(x_i, x_j) \ \prod_{l \neq i} q(\sigma(x_l))  \
\le
\end{equation}
$$
c K_{n-1}  e^{t ( L^i_{max} + L^j_{max})} \alpha(\tau(x_i) -
\tau(x_j)) \  \prod_{l=1}^n  q(\sigma(x_l)).
$$
Using formula (\ref{25}), the Fourier transform and the Fubini
theorem we finally obtain from (\ref{aq}) - (\ref{36}) the upper
bound on $v^{(n)}_{i,j}$:
$$
v^{(n)}_{i,j} (x_1, \ldots, x_n) \ = \ \int_0^{\infty} e^{t \hat L_n^{\ast}}  f(x_1, \ldots, x_n) \ dt \
\le
$$
$$
c K_{n-1} \prod_{i=1}^n q(\sigma(x_i)) \left| \int_0^{\infty}
\int_{R^d} e^{t (\hat \alpha (p) + \hat \alpha (-p) -2)} \hat
\alpha(p) dp dt \right| \ = \   c A K_{n-1} \prod_{i=1}^n
q(\sigma(x_i)),
$$
where
\begin{equation}\label{A}
A = \frac{1}{(2 \pi)^d}\left| \int_0^{\infty} \int_{R^d \backslash
\{0\} } e^{t (\hat \alpha (p) + \hat \alpha (-p) -2)} \hat \alpha(p)
dp dt \right| \le \int_{R^d} \frac{|\hat \alpha(p)|}{2- \hat \alpha
(p) - \hat \alpha (-p)} \ dp < \infty
\end{equation}
when $d \ge 3$.

Since the function $f^{(n)}$ is the sum
of $n(n-1)$ similar terms, then the function $ k^{(n)}$
given by (\ref{sumv}) is bounded by the function
$$
C n^2 K_{n-1} \prod_{i=1}^{n-1} q(\sigma(x_i))
$$
for some $C>0$. Thus we get the recurrence inequality
\begin{equation}\label{50}
K_n \ \le \ C n^2 K_{n-1},
\end{equation}
which is valid under
\begin{equation}\label{49}
K_n \ = \ C^n \ (n!)^2.
\end{equation}
Thus
\begin{equation}\label{est43}
k^{(n)} (x_1, \ldots, x_n) \ \le \  C^n (n!)^2 \prod_{i=1}^n q(\sigma(x_i)),
\end{equation}
where $ k^{(n)}$ is defined by (\ref{sumv}).

Moreover, using the positivity of $f(x_1, \ldots, x_n)$ (see (\ref{32})), inequality (\ref{36}), the Fourier transform and the Fubini theorem as above, we get from (\ref{35})
$$
v^{(n)}_{i,j}(x_1, \ldots, x_n)  \ =  \ \int_0^{\infty} \left( e^{t \hat L_n^{\ast}} f \right) (x_1, \ldots, x_n) dt \ \le
$$
$$
\frac{1}{(2 \pi)^d} c K_{n-1} \prod_{i=1}^n q(\sigma(x_i))
\int_{R^d}  \frac{\hat \alpha(p) \ e^{-ip(\tau(x_i) - \tau(x_j)) }}{2- \hat \alpha
(p) - \hat \alpha (-p)} \ dp.
$$
Integrability of the function $\frac{|\hat \alpha(p)|}{2- \hat
\alpha (p) - \hat \alpha (-p)}$ (see (\ref{A})) implies by the
Lebesgue-Riemann lemma that the function $v^{(n)}_{i,j}$ satisfies the following condition:
\begin{equation}\label{38A}
v^{(n)}_{i,j}(x_1, \ldots, x_n) \ \to \ 0 \quad \mbox{ when } \;
|\tau(x_i) - \tau(x_j)|  \to \infty.
\end{equation}
Consequently, using (\ref{sumv}) and (\ref{38A}) we conclude that
\begin{equation}\label{100}
\left( - \hat L_n^{\ast} \right)^{-1} f^{(n)} (x_1, \ldots, x_n)  \ = \
\sum_{i \neq j} v^{(n)}_{i,j}  (x_1, \ldots, x_n) \ \to \ 0,
\end{equation}
when $|\tau(x_i) - \tau(x_j)|  \to \infty$ for all $i \neq j$.
Thus we constructed $\left( -\hat L_n^{\ast} \right)^{-1} f^{(n)}$ meeting estimate (\ref{est43}) and condition (\ref{100}).
\\

For a given $n$ the equation  (\ref{Last}) is an inhomogeneous linear equation. Then the general solution $ k^{(n)} (x_1, \ldots, x_n) $ of (\ref{Last}) has the form
$$
k^{(n)} (x_1, \ldots, x_n) \ = \ \int_0^{\infty} e^{t \hat
L_n^{\ast}} f^{(n)} (x_1, \ldots, x_n) \ dt \ +  \ A_n \
\prod_{i=1}^{n} q(\sigma(x_i)),
$$
where $A_n$ are some constants. If we are looking for the set of
correlation functions $k_\varrho^{(n)}$ for which
\begin{equation}\label{51}
| k_\varrho^{(n)} (x_1, \ldots, x_n) \ - \ \varrho^n \prod_{i=1}^{n}
q(\sigma(x_i))| \ \to \ 0,
\end{equation}
when  $|\tau(x_i) - \tau(x_j)| \to \infty$  for all $i\neq j$, then taking into account (\ref{100}) we put
\begin{equation}\label{52}
k^{(n)}_\varrho (x_1, \ldots, x_n) \ = \  \int_0^{\infty} e^{t \hat L_n^{\ast}} f^{(n)} (x_1, \ldots, x_n)
dt \ + \ \varrho^n \prod_{i=1}^{n} q(\sigma(x_i)).
\end{equation}
It is clear that the last term in (\ref{52}) vanishes under the
action of $\hat L_n^{\ast}$, and (\ref{51}) holds because we got
(\ref{100}).

In this case instead of (\ref{50}) we have the recurrence
\begin{equation}\label{53}
K_n \ \le \ C n^2 K_{n-1} \ + \ \varrho^n.
\end{equation}
Taking $L_n=\frac{K_n}{C^n (n!)^2}$ we have
$$
L_n  \ \le \ L_{n-1} \ + \ \frac{\varrho^n}{C^n (n!)^2} \ \le \ D
$$
with some positive constant $D>0$, and we take
\begin{equation}\label{55}
K_n \ = \ D C^n (n!)^2,
\end{equation}
which differs from (\ref{49}) only by the constant factor, and the estimate (\ref{est}) is proved.

Thus we proved the existence of solutions $\{ k_{\varrho}^{(n)} \}$
of the system (\ref{Last}) corresponding to the stationary problem.
To verify that this system of correlation function is associated
with a measure $\mu_\varrho$ on the configuration space, we will prove
in the next section that the measure $\mu_\varrho$ can be constructed
as a limit of an evolution of measures $\mu_\varrho^{(t)}$
associated with the solutions of the Cauchy problem (\ref{59}) with
corresponding initial data (\ref{60}).

\section{The proof of Theorem 1. The Cauchy problem.  }

In this section we find the solution of the Cauchy problem
(\ref{59}) - (\ref{60}) and prove the convergence (\ref{Th1-2}).
Using Duhamel formula we have
\begin{equation}\label{61}
k^{(n)}(t) \ = \  k^{(n)}(0) \ + \  \int_0^t e^{(t-s) \hat
L_n^{\ast}} f^{(n)}(s) \ ds,
\end{equation}
where $f^{(n)}(s)$ is expressed through $k^{(n-1)}(s)$ via
(\ref{f}). We also used there that the operator  $\hat L_n^{\ast}$
annihilates $k^{(n)}(0)$ of the form (\ref{60}).

Let us notice that $k_\varrho^{(n)}$ given by (\ref{52}) have no
product form (\ref{60}). We have
$$
k^{(n)}(t) -  k_\varrho^{(n)} \ = \ \left( e^{t \hat L_n^{\ast}} - 1
\right) k_\varrho^{(n)} \ +
$$
$$
e^{t \hat L_n^{\ast}}(k^{(n)}(0) - k_\varrho^{(n)}) \ + \  \int_0^t
e^{(t-s) \hat L_n^{\ast}} f^{(n)}(s) \ ds \ =
$$
\begin{equation}\label{64}
e^{t \hat L_n^{\ast}}(k^{(n)}(0) - k_\varrho^{(n)}) \ + \ \int_0^t
e^{(t-s) \hat L_n^{\ast}} (f^{(n)}(s) -  f_\varrho^{(n)}) \ ds.
\end{equation}
Here $f_\varrho^{(n)}$ are expressed in terms of $k_\varrho^{(n-1)}$
by (\ref{f}), and we used that the equation $ \hat L_n^{\ast}
k_\varrho^{(n)} \ = \ - f_\varrho^{(n)}$ implies
$$
\left( e^{t \hat L_n^{\ast}} - E \right) k_\varrho^{(n)} \ = \ -
\int_0^t \frac{d}{ds} e^{(t-s) \hat L_n^{\ast}} k_\varrho^{(n)} ds \
\ = \ - \int_0^t e^{(t-s) \hat L_n^{\ast}}  f_\varrho^{(n)} \ ds
$$
We shall prove now that both terms in (\ref{64}) converge to 0 in
sup-norm of $X_n$.

For the first term using inversion formula (\ref{52}) and (\ref{60})
we have
\begin{equation}\label{65}
e^{t \hat L_n^{\ast}}(k^{(n)}(0) - k_\varrho^{(n)}) \ = \ e^{t \hat
L_n^{\ast}}(k^{(n)}(0) - v^{(n)} -  k^{(n)}(0))  \ = \ - e^{t \hat
L_n^{\ast}}v^{(n)},
\end{equation}
where
\begin{equation}\label{66}
v^{(n)} \ = \ \int_0^{\infty} e^{s \hat L_n^{\ast}} f_\varrho^{(n)}
\ ds.
\end{equation}
Since
$$
f_\varrho^{(n)}(x_1, \ldots, x_n) \ = \ \sum_{i,j:\ i\neq j}
k_\varrho^{(n-1)}(x_1, \ldots,\check{x_i}, \ldots, x_n) \ a(x_i,
x_j),
$$
then
$$
v^{(n)}(x_1, \ldots, x_n) \ = \ \sum_{i,j:\ i\neq j} \int_0^{\infty}
 e^{s \hat L_n^{\ast}} k_\varrho^{(n-1)}(x_1, \ldots,\check{x_i},
\ldots, x_n) \ a(x_i, x_j) \ ds.
$$
To prove that $\| e^{t \hat L_n^{\ast}}v^{(n)} \|_{X_n} \to 0$ as $t
\to \infty$ it is enough to prove that its Fourier transform tends
to 0 in $L^1$ norm when $t\to\infty$.

Using the estimate (\ref{estimate}) on $k^{(n)}_\varrho$ together
with the inequality (\ref{aq}) on $a(x_i, x_j)$ we can estimate
$e^{t \hat L_n^{\ast}} v^{(n)}$ applying the monotonicity of $ e^{t
\hat L_n^{\ast}}$ and (\ref{24}) - (\ref{25}):
$$
\left| \left( e^{t \hat L_n^{\ast}} \ v^{(n)} \right) (x_1, \ldots,
x_n) \right| \ \le
$$
$$
D C^{n-1} ((n-1)!)^2  e^{t \hat L_n^{\ast}} \sum_{i\neq j}
\int_0^{\infty} e^{s \sum_{i=1}^n L^{i}} c \prod_{i=1}^n
q(\sigma(x_i)) \ \alpha(\tau(x_i) - \tau(x_j)) \
 ds \ \le
$$
$$
D C^{n-1} ((n-1)!)^2 c  \prod_{i=1}^n q(\sigma(x_i)) \ \sum_{i\neq
j} \int_{R^{2d}}  e^{t (\hat \alpha(p_i) + \hat \alpha (p_j) -2)}
$$
$$
\int_0^{\infty} e^{s (\hat \alpha(p_i) + \hat \alpha (p_j) -2)} \
 |\hat \alpha (p_i)| \ \delta(p_i+p_j) \ ds dp_i dp_j \ \le
$$
$$
D C^{n-1} (n!)^2 c  \prod_{i=1}^n q(\sigma(x_i)) \int_{R^d} e^{t
(\hat \alpha(p) + \hat \alpha (-p) -2)} \frac{|\hat \alpha (p)|}{2 -
\hat \alpha(p) - \hat \alpha (- p)} \ dp.
$$
Here the presence of $\delta$-function corresponds to the shift
invariance. Since the function $\frac{|\hat \alpha (p)|}{2 - \hat
\alpha(p) - \hat \alpha (- p)} $ is integrable in the momentum space
for $d \ge 3$ and $ \hat \alpha(p) + \hat \alpha (- p) <2 $ for $p
\neq 0$, then the function
$$
\tilde A \  e^{t(\hat \alpha(p) + \hat \alpha (-p) -2)} \
\frac{|\hat \alpha (p)|}{2 - \hat \alpha(p) - \hat \alpha (- p)}.
$$
tends to 0 in $L^1$ norm (in "momentum" variables $p$) when $t \to
\infty$. Consequently its inverse Fourier transform tends to 0 in
$X_n$ norm (i.e. in sup-norm) when $t \to \infty$. Thus we proved
that the first term in (\ref{64}) tends to 0 in sup-norm when $t \to
\infty$.

We consider now the second term in (\ref{64}) and will prove that
\begin{equation}\label{81}
\int_0^t e^{(t-s) \hat L_n^{\ast}} (f^{(n)}(s) - f_\varrho^{(n)}) \
ds \ \to 0
\end{equation}
in sup-norm when $t \to \infty$ using induction assumption that
\begin{equation}\label{ind}
\| k^{(n-1)}(t) \ - \  k_\varrho^{(n-1)} \|_{X_{n-1}} \ \to \ 0
\quad \mbox{ as} \; t \to \infty.
\end{equation}
As the first step of induction we have
\begin{equation}\label{82}
k^{(1)}(t,x) \ \equiv \ k_\varrho^{(1)}(x) \ = \ \varrho \
q(\sigma(x)).
\end{equation}
Next by induction assumption (\ref{ind}) implies that
\begin{equation}\label{83}
\| k^{(n-1)}(t) \|_{X_{n-1}}  \ \le \  M_{n-1} \quad \mbox{ for all
} \; t \ge 0
\end{equation}
with some positive constant depending only on $n$. Indeed, the
operator $\hat L_n^{\ast}$ is bounded and the function $a(x,y)$ is
bounded, hence the norm of the solution $k^{(n)}$ of the problem
(\ref{59}) (with initial data uniformly bounded for $l \le n$) is
evidently bounded on any compact time interval $[0,\tau]$. On the
other hand, for any $\varepsilon >0$ there exists $\tau$ such that
for all $t > \tau$ the norm $\| k^{(n-1)}(t) - k_\varrho^{(n-1)}\|
<\varepsilon$ by (\ref{ind}). Thus the bound (\ref{83}) is proved.

From (\ref{ind}) it follows that
\begin{equation}\label{83A}
\| f^{(n)}(t) \ - \  f_\varrho^{(n)}\|_{X_n} \ \to 0 \quad \mbox{
as} \; t \to \infty.
\end{equation}
To estimate the integral (\ref{81}) we split the integral as follows
\begin{equation}\label{84}
\left( \int_0^\tau \ + \ \int_\tau^t \right) e^{ s \hat L_n^{\ast}}
(f^{(n)}(t-s) - f_\varrho^{(n)}) \ ds.
\end{equation}

Let us estimate the second integral in (\ref{84}) using the
monotonicity of the semigroup $ e^{ s \hat L_n^{\ast}}$:
\begin{equation}\label{85}
\left| \int_\tau^t  e^{s \hat L_n^{\ast}} (f^{(n)}(t-s) -
f_\varrho^{(n)}) \ ds \right| \  \le \  \int_\tau^t  e^{s \hat
L_n^{\ast}} \left( |f^{(n)}(t-s)| + |f_\varrho^{(n)}| \right) \ ds \
\le
\end{equation}
$$
\left( M_{n-1} \ + \ \| k_\varrho^{(n-1)} \| \right)  \int_\tau^t
e^{s \hat L_n^{\ast}} \sum_{i \neq j} a(x_i, x_j) \ ds.
$$
Using the inequality analogous to (\ref{aq})
$$
a(x_i, x_j) \ \le \ \tilde c  q(\sigma(x_i)) q(\sigma(x_j)) \alpha
(\tau(x_i) - \tau(x_j))
$$
we conclude that it will be sufficient to estimate for any pair $i
\neq j$ the following integral
\begin{equation}\label{89}
\int_\tau^{t} \int_{R^d} e^{s(\hat \alpha(p) + \hat \alpha (-p) -2)}
|\hat \alpha (p)| \ dp \ ds \ \le \  \int_\tau^{\infty} \int_{R^d}
e^{s(\hat \alpha(p) + \hat \alpha (-p) -2)} |\hat \alpha (p)| \ dp \
ds
\end{equation}

Since the integral
\begin{equation}\label{90}
\int_0^{\infty} \int_{R^d} e^{s(\hat \alpha(p) + \hat \alpha (-p)
-2)} |\hat \alpha (p)| \ dp \ ds \ = \ \int_{R^d} \frac{|\hat \alpha
(p)|}{2 - \hat \alpha(p) - \hat \alpha (- p)} \ dp
\end{equation}
converges, then the integral (\ref{89}) tends to 0 when $\tau \to
\infty$. Consequently we can take $\tau$ in such a way that
(\ref{89}) is less than $\varepsilon$, and then (\ref{85}) is less
than $C \varepsilon$ for some $C$ and any $t>\tau$.

Finally let us estimate the first integral in (\ref{84}) for a given
$\tau$:
\begin{equation}\label{91}
\int_0^\tau  e^{ s \hat L_n^{\ast}} (f^{(n)}(t-s) - f_\varrho^{(n)})
\ ds.
\end{equation}
From (\ref{83A}) it follows that we can choose $t_0>\tau$ such that
for $t> t_0$ the following estimate holds
$$
\| f^{(n)}(t-\tau) - f_\varrho^{(n)} \|_{X_n} \ < \
\frac{\varepsilon}{\tau}.
$$
Consequently the norm of (\ref{91}) is less than $\varepsilon$.
Finally, for $t>t_0$ the integral in (\ref{81}) is less than
$(C+1)\varepsilon$ in sup-norm and convergence to zero of (\ref{81}) as well as of
(\ref{64}) is proved.

Thus we proved the strong convergence (\ref{Th1-2}). Using results
from \cite{KS} we can conclude that the solution $\{ k_\varrho^{(n)}(t) \}$ of the Cauchy problem (\ref{59}) is a system
of correlation functions corresponding to the evolution of states
$\{ \mu_t \}$. The construction of the measure $\mu_\varrho$ from the family of correlations functions
$$
k^{(n)}_\varrho \ = \  \lim_{t\to\infty} k^{(n)}_\varrho (t),
$$
where $k^{(n)}_\varrho$ is a solution (\ref{52}) of the system
(\ref{Last}), is based on the Lenard positivity of this family, see \cite{KKP}.

\section {Concluding remarks and the proof of Theorem 2.}

{\bf Remark 1.} {\it If instead of (\ref{60}) we take the initial
data in the form
\begin{equation}\label{93}
\bar k^{(n)}|_{t=0} (x_1, \ldots, x_n) \ = \ \varrho^n \
\prod_{i=1}^{n} h(\sigma(x_i))
\end{equation}
for some $ \varrho >0$ and some normalized positive function $h(s),
\; \int_S h (s) d\nu(s)=1$ (not necessarily the eigenfunction of $Q$
and not necessarily continuous), then we have convergence
$$
\bar k^{(n)}(t) \  \to \  k^{(n)}_{\varrho_1}, \quad t \to \infty,
$$
where $k^{(n)}_{\varrho_1}$ is defined by the formula (\ref{52}),
and
\begin{equation}\label{97}
\varrho_1  \ = \ \frac{\varrho \langle h, \ \tilde q
\rangle}{\langle q, \ \tilde q \rangle} \ = \ \frac{ \varrho \int_S
h(s) \ \tilde q (s)\ d\nu(s)}{\int_S q(s) \ \tilde q (s)\ d\nu(s)}.
\end{equation}
Here $\tilde q$ is the positive eigenfunction of the adjoint
operator $Q^{\ast}$. }

To prove this convergence we should make some small modifications in
the above reasoning.

1) In (\ref{65}) we have the additional term
$$
e^{ t \hat L_n^{\ast}} (\bar k^{(n)}(0) - k^{(n)}(0, \varrho_1)).
$$
This term does not depend on space coordinates and equals to
\begin{equation}\label{96}
\otimes_i  e^{ t (Q^i - E)} (\bar k^{(n)}(0) - k^{(n)}(0,
\varrho_1)).
\end{equation}
Here $Q^i$ is the operator $Q$ acting on $i$-th spin variable. From
the Krein-Rutman theorem it follows that (\ref{96}) tends to 0 if
(\ref{97}) is fulfilled.

2) The condition (\ref{82}) is also violated. The first correlation
function $k^{(1)}(t)$ now depends on time ( but not depends on space
variables) and satisfies an equation
$$
\frac{\partial k^{(1)}}{\partial t } \ = \ \hat L_1^{\ast} \ k^{(1)}
\ = \ (Q-E) \ k^{(1)}
$$
with the initial data $\bar k^{(1)}|_{t=0}= \varrho \ h(s)$. Then
$$
 k^{(1)} (t) \ \to \ \varrho_1 \ q(s), \quad t \to \infty
$$
by the Krein-Rutman theorem provided the density equation (\ref{97})
is fulfilled.
\\

The same approach can be applied when the initial data are mixtures
of marked Poisson fields with different spatial densities and
different mark distributions (provided marks are mutually
independent).
\\

{\bf Remark 2. Law of large numbers.} {\it Theorem 1 implies the
correlation decay
$$
| k^{(2)}_\varrho (x_1, x_2) \ - \ k^{(1)}_\varrho (x_1)
k^{(1)}_\varrho (x_2) | \ \to \ 0,
$$
when  $|\tau(x_1) - \tau(x_2)| \to \infty$. By the standart
application of the Chebyshev inequality, see e.g. \cite{R}, we get
the low of large numbers for the number of particles, i.e. the
existence of the spatial density of particles: }
$$
\frac{N(V)}{V} \ \to \ \varrho, \quad \mbox{ as } \ V \to \infty.
$$
\\

{\bf The proof of Theorem 2 } is completely analogous to the proof
of Theorem 1, when $m(s) \equiv 1$. We only should check that
$$
\tilde L_n^{\ast}  \prod_{i=1}^{n} g(\sigma(x_i)) \ = \ 0.
$$
Indeed, we have
$$
\tilde L_n^{\ast}  \ = \ \sum_{i=1}^n  \tilde L^i,
$$
$$
\tilde L^i \ \prod_{j=1}^{n} g(\sigma(x_j)) \ =
$$
$$
\prod_{j \neq i} g(\sigma(x_j)) \left( -m(\sigma(x_i))
g(\sigma(x_i)) + \kappa_{cr} \int_S Q(\sigma(x_i), s') g(s') d\nu
(s') \right) \ = \ 0,
$$
since $\kappa_{cr} \tilde Q g \ = \ g$ implies
$$
\kappa_{cr} \int_S Q(s, s') g(s') d\nu (s') \ = \ m(s) \ g(s).
$$
Theorem 2 is proved.
\\

{\bf Acknowledgements.} We thank the anonymous referees for helpful remarks.

\end{document}